\documentclass[aps,pra,preprint,superscriptaddress]{revtex4}
\usepackage{amsmath}
\usepackage{gensymb}
\usepackage{subfig}
\usepackage{graphicx}
\usepackage{float}
\usepackage{amsthm}
\usepackage{amssymb}
\usepackage{xfrac}
\usepackage[toc,page]{appendix}
\usepackage{setspace}
\usepackage{lscape}
\usepackage[utf8]{inputenc}
\usepackage{tikz}

\newcommand{\half}{\frac{1}{2}}
\newcommand{\gm}{\gamma}

\newcommand{\bt}{\beta}
\newcommand{\alp}{\alpha}

\newcommand{\non}{\nonumber}

\newcommand{\Gm}{\Gamma}

\newcommand{\zt}{\zeta}
\newcommand{\sbp}{\subparagraph*{}}
\newcommand{\comment}[1]{}

\newcommand{\bds}{\boldsymbol}

\newcommand{\alld}{\allowdisplaybreaks}

\newcommand{\eqtion}[1]{\begin{equation} \non #1 \end{equation}}

\usepackage{hyperref}

\begin{document}
\title{Quantum Secrecy in Thermal States III}
\author{Anne Ghesqui\`ere}
\author{Benjamin T. H. Varcoe}
\affiliation{Quantum Experimental Group, School of Physics and Astronomy, University of Leeds, Leeds LS2 9JT, United Kingdom}
\email[]{b.varcoe@leeds.ac.uk}

\date{\today}

\begin{abstract}
In this paper we expanded the security of a central broadcast protocol using thermal states to the case in which the eavesdropper controls the source.
Quantum secrecy in a continuous variable central broadcast scheme is guaranteed by the quantum correlations present in thermal states arising from the Hanbury Brown and Twiss effect.
This work allows for a method of key exchange in which two parties can agree a key as long as both can detect the same source and they are within the spatial coherence length of the source.
This is important because it allows quantum secure key exchange with only minimal changes to existing infrastructure. 
\end{abstract}

\maketitle


Expanding Quantum Key Distribution outside of the normal paradigm of single photons sent from Alice to Bob is a current challenge that is particularly relevant to traditional networks where upgrading to single photons optical fibers is impractical. 
In a series of papers \cite{sakuya:2017_1, sakuya:2018_1}, we have been developping a model for thermal states key distribution that is especially relevant to this challenge because it would allow QKD in a microwave network.
In these papers we demonstrate that, rather than being a source of `noise'\cite{Weedbrookpra:2012}, thermal states are a source of discord that can be used as a resource for quantum computing\cite{Pirandola:2014}.

In particular the model that we consider is a central broadcast \cite{Maurer:1999} model rather than a point-to-point model \cite{BB:84}. 
In central broadcasting, a single source transmits a signal to several receivers, which allows the receivers to detect Hanbury Brown and Twiss noise correlations.
A common example of a central broadcast channel is that of a satellite beaming in free-space to several receiving antennas and the information received depends on the channel noise and the reliability of the detectors. 

This set-up is easy to implement in the laboratory, using the Hanbury Brown and Twiss (HBT) interferometer \cite{HBT:1956_1, HBT:1956_2}.
A source shines onto a beamsplitter, which divides the signal into two beams, one detected by Alice, the other by Bob.
When the radiation is thermal, the signal is composed of bunched pairs.
The action of the beamsplitter is to split the pairs, to exploit the correlations within.
This is known as the Hanbury Brown and Twiss effect, and the correlations are known to exhibit discord \cite{ragy:2013}, which is a necessary condition for QKD \cite{Pirandola:2014}.








Therefore, far from being dismissable as mere noise \cite{Weedbrookpra:2012}, the discord present in thermal photons serves as a resource in QKD. 
In ref.  \cite{sakuya:2017_1, sakuya:2018_1} we have described a central broadcast scheme (CBS) where a source (controlled by Alice) emits a signal which is divided and shared between the two legal parties. 
There are two channels open to the eavesdropper in such a protocol, the lower channel between the beamsplitter and Bob, and the higher channel between the source and the beamsplitter.
In \cite{sakuya:2017_1}, we let Eve access the lower channel; in \cite{sakuya:2018_1}, we gave her access to the higher channel.
In both cases, the secrecy arises from the correlations present in the bunched pairs.

Only one step remains to fully establish our CBS as a valid quantum key distribution, and that is the requirement of trust in the source.
The security of the source is generally considered paramount to quantum key distribution protocols.
Eavesdropping is allowed on the way from source to the legal parties, but the source cannot be untrusted.



In a CBS, the source can be either under (for instance) Alice's control, or out of either legal parties' control.
So far, we have considered that it was under Alice's control.
This has allowed us to make the assumption that the radiation transmitted to Alice and Bob is thermal. 
Technically, no such assumption can be made if the source is out of Alice's or Bob's control.
It is natural to wonder then, if the protocol remains secure in this case.

\begin{figure}[h]
\centering
\includegraphics[scale=0.45]{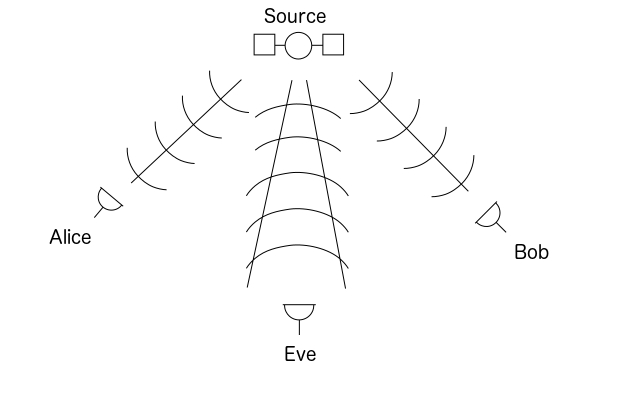}
\caption{In this situation, a satellite beams down a signal, which is received by Alice and Bob. We assume that Eve has control of the satellite, and so beams up the signal she wants transmitted to Alice and Bob.}
\label{mistresseve_satsetup}
\end{figure}

Let us recall the protocol briefly, as illustrated on Figure~\ref{mistresseve_satsetup}.
A source produces states which are sent onto a beamsplitter $\eta_{ab}$, which splits it into one part which goes to Alice and one part that goes to Bob.
Alice and Bob measures their signal separately.
They perform a $g^{(2)}(0) > 1$ check on their data, to ensure its thermality and therefore ascertain the presence of correlations.
Once the presence of correlations is confirmed, they proceed as per usual, with reconciliation and privacy amplification.

The $g^{(2)}(0)$-check is the pivot here, because it is when Alice and Bob verify that the signal they have detected is indeed, correlated. 
If it is not, they must begin anew.
This limits the eavesdropper in her actions, and forces her to emit a thermal signal.
If she wants to beat the $g^{(2)}(0)$-check and retain some information about the signal which Alice and Bob will detect, Eve must create a signal which will correlate her to both.
Thankfully, the likelyhood of this is very remote for three reasons, firstly because three-way correlations, where the bunched radiation is not a pair but a triplet, is a statistically rare occurence.
Secondly,  Eve would face the time limitation introduced by the coherence length of the radiation.
Simply put, Eve must somehow find herself in a physical position where she can force the satellite to produce a sufficient number of triplets for her to be correlated to Alice and to Bob, and as well, detect her share of the signal within the coherence time. 
Lastly, the splitting at the beamsplitter is stochastic and therefore Eve cannot predict it.

This is not without caveats for Alice and Bob.
Since they can have no trust in the source, they must now reconcile against her, at the cost of a lower key rates. 
We are considering here, a retrofit to a broadcast channel.
Our aim is to build trust in existing infrastructures, where the origin of the data is uncertain, by establishing that it can be used to distribute a quantum secure secret key, on the condition that the  $g^{(2)}(0)$ requirement is satisfied. 


In the following, we model the eavesdropping and establish the security of the protocol for correlated and un-correlated noise.

\begin{figure}[h]
\centering
\includegraphics[scale=0.45]{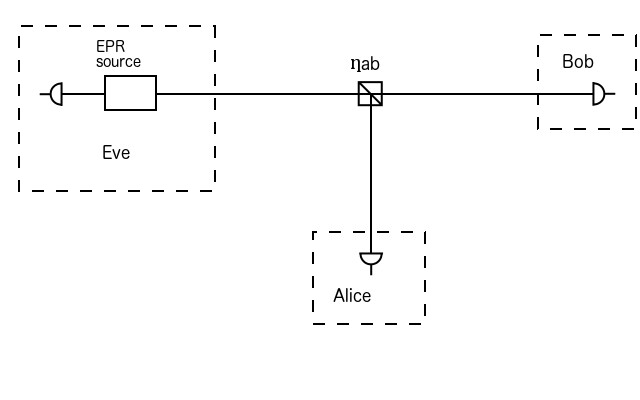}
\caption{Schematics of the set-up. A thermal source shines onto a beamsplitter $\eta_{ab}$ which splits the signal into a part going to Alice and the other going to Bob. We assume that Eve has control of the source. The EPR is split into one mode which is the source signal for Alice and Bob, the other remains Eve's. }
\label{theosetup}
\end{figure}

\section{Formal description}

\sbp To build our model, we follow Figure~\ref{theosetup} and consider that Eve is sending one mode of an EPR state down.
This is the equivalent of a prepare-and-send scheme, but instantly it puts Eve at a disadvantage since upon her measurement of her mode, the mode sent to the legal parties is modelled by a thermal state \cite{Weedbrookrmp:2012}.
That mode falls onto a beamsplitter with transmittance $\eta_{ab}$, which divides it between Alice and Bob. 
Eve's input state is \eqtion{\Gm_{EPR} = \left(\begin{array}{cc} \nu \bds{I}_2 & \sqrt{\nu^2-1}\bds{Z}_2 \\  \sqrt{\nu^2-1}\bds{Z}_2 & \nu \bds{I}_2\end{array} \right)\,,} 
where $\bds{I}_2$ is the $2\times2$ identity matrix and $\bds{Z}_2 = \left( \begin{array}{cc}1 & 0 \\0 & -1  \end{array}\right)$. 

We assume a vacuum at the other input of the beamsplitter. 
The outcome is (with $\zt = \sqrt{\nu^2 -1}$ and $\mu_{ab} = \sqrt{1-\eta_{ab}}$ for clarity)
\eqtion{\Gm = \left( \begin{array}{cccccc} \nu & 0 & \sqrt{\eta_{ab}} \zt & 0  & -\mu_{ab} \zt & 0 
\\ 0 & \nu  & 0 & -\sqrt{\eta_{ab}} \zt  & 0 &  \mu_{ab} \zt
\\ \sqrt{\eta_{ab}} \zt  & 0 &  \eta_{ab} \nu + \mu_{ab}^2 & 0 & \mu_{ab} \sqrt{\eta_{ab}} (1-\nu) & 0 
\\  0 & - \sqrt{\eta_{ab}} \zt & 0 & \eta_{ab} \nu + \mu_{ab}^2 & 0 & \mu_{ab} \sqrt{\eta_{ab}} (1-\nu) 
\\ -\mu_{ab} \zt & 0 & \mu_{ab} \sqrt{\eta_{ab}} (1-\nu) & 0 & \mu_{ab}^2 \nu + \eta_{ab} & 0 
\\  0 & \mu_{ab} \zt & 0 & \mu_{ab} \sqrt{\eta_{ab}} (1-\nu) & 0 & \mu_{ab}^2 \nu + \eta_{ab} \end{array} \right)\,.} 
We identify the blocks as \eqtion{\Gm = \left( \begin{array}{ccc} \gm_e & \gm_{eb} & \gm_{ea} \\ \gm_{eb} & \gm_b & \gm_{ab} \\ \gm_{ea} & \gm_{ab} & \gm_a \end{array}\right)\,.}
As we mentioned before, since the eavesdropper controls the source, the secrecy is determined by the information the legal parties share independently of the source, $I(A:B|S)$, which is defined as
\eqtion{I(A:B|S) = H(a,s) + H(b,s) - H(s) - H(a,b,s) \,.}
To calculate $H(a,s)$, $H(b,s)$, $H(s)$ and $H(a,b,s)$, we need 
\begin{align}
\Gm_{as} =& \left( \begin{array}{cc} \gm_e & \gm_{ea} \\ \gm_{ea} & \gm_a \end{array} \right) \non
\\ \Gm_{bs} =& \left( \begin{array}{cc} \gm_e & \gm_{eb} \\ \gm_{eb} & \gm_b \end{array} \right) \non
\\ \Gm_s =& \gm_e \non
\\ \Gm_{abs} =& \left( \begin{array}{ccc} \gm_e & \gm_{eb} & \gm_{ea} \\ \gm_{eb} & \gm_e & \gm_{ab} \\ \gm_{ea} & \gm_{ab} & \gm_a \end{array} \right) \non
\end{align}
respectively. 
Then \cite{Shannon:1948_3, GarciaPatron:07},
\eqtion{I(A:B|S) = \half \log (2\pi e)^2 \det(\Gm_{as}) + \half \log (2\pi e)^2 \det(\Gm_{bs}) - \half \log (2\pi e) \det(\Gm_s) - \half \log (2\pi e)^3 \det(\Gm_{abs})  \,. }

Alice and Bob expect correlations in their respective signals. 
These correlations are quantified using the quantum discord, defined explicitly as \eqtion{D(B|A) = S(\Gm_a) - S(\Gm_{ab}) +\min_{\Gm_0} S(\Gm_{b|x_A})\,}
where $\Gm_{b|x_A}$ is the covariance matrix of B conditioned by a homodyne measurement on A \cite{Weedbrookrmp:2012} \eqtion{\Gm_{b|x_A} = \, \Gm_b - \Gm_{ab}(X\Gm_a X)^{-1} \Gm_{ab}^T\,,} 
with $X = \left(\begin{array}{cc} 1& 0 \\ 0 & 0 \end{array}\right)$ and $()^{-1}$ the pseudo-inverse.
The Von Neumann entropy is given by \eqtion{S(x) = \sum_{i=1}^N \left(\frac{x_i+1}{2}\right)\log\left(\frac{x_i+1}{2}\right) - \left(\frac{x_i-1}{2}\right)\log\left(\frac{x_i-1}{2}\right)} 
where $x_i$ are the symplectic eigenvalues of $\Gm$.

\section{Influence of correlated noise in Eve's channel}
\subsection{Coherent state}
\begin{figure}[h]
\centering
\includegraphics[scale=0.35]{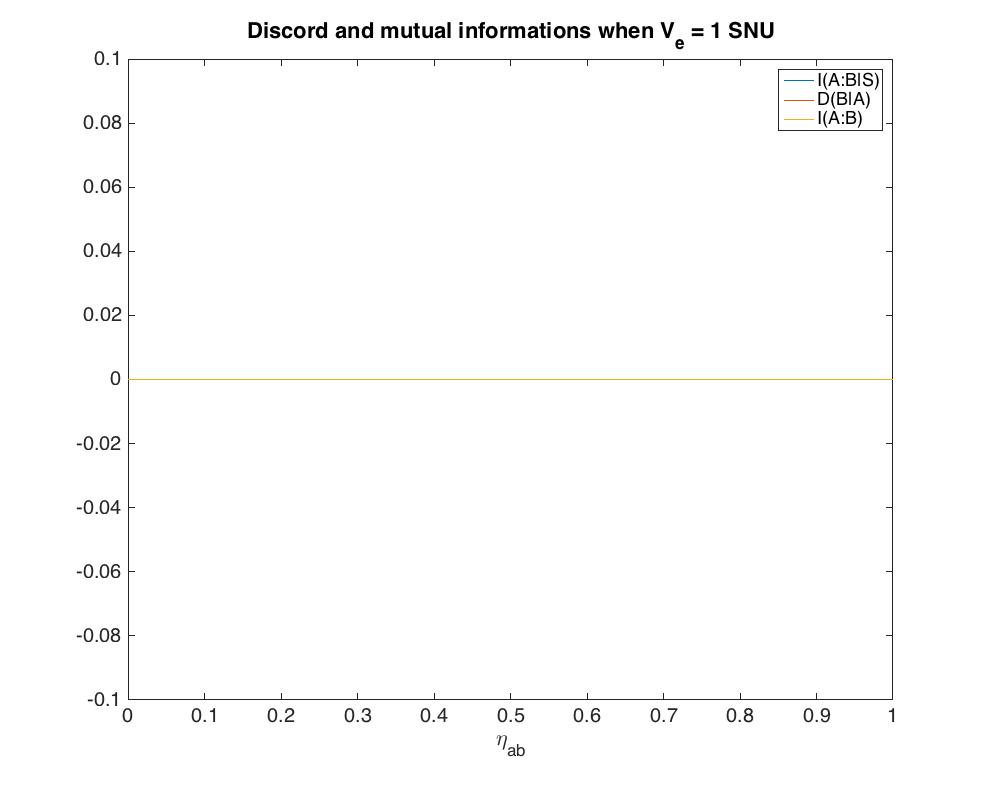}
\caption{We plot the discord (red) and conditional mutual information (blue) against $\eta_{ab}$ for $V_e = 1SNU$.}
\label{cohEve}
\end{figure}

\sbp $V_e = 1 SNU$ (Shot Noise Unit) represents a vacuum state, which is the minimum uncertainty state, centred at the origin.
Because a coherent state is a displaced vacuum state, its variance is also $V_e =1 SNU$. 
We have established before in \cite{sakuya:2017_1, sakuya:2018_1} that there are no correlations, nor information in a coherent state when it is shared by central broadcast. 
Figure~\ref{cohEve} confirms this. 

It would be incorrect to assume that this  is a general model for a system running on empty as it were. 
We consider here loss-less and noise-less channels between Eve and $\eta_{ab}$, as well as between $\eta_{ab}$ and Alice and Bob.
Therefore, we can establish the amount of correlations which Alice and Bob can expect, should they share a coherent state. 
For this reason, we have also elected to plot not only $I(A:B|S)$ but also $I(A:B)$; this shows that there is no mutual information between Alice and Bob for them to reconcile against the source. 
This is confirmed by the nullity of the discord, which demonstrates that there are no correlations.

\subsection{Thermal state}

Eve inputs a state $V_e = \nu + 1$, where $\nu$ is the variance of the thermal state she sends through the satellite. 
\begin{figure}[h]
\centering
\includegraphics[scale=0.35]{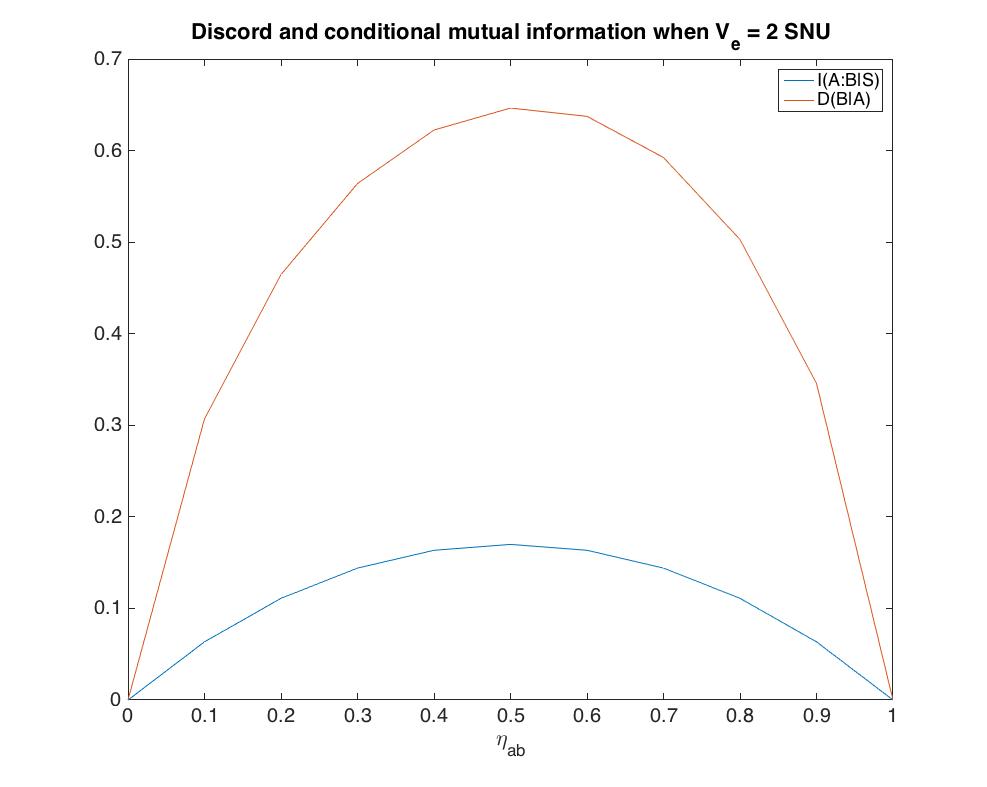}
\caption{We plot the discord (red) and conditional mutual information (blue) against $\eta_{ab}$ for $V_e = 2SNU$.}
\label{cohEveSN}
\end{figure}

This situation from Eve's point of view, should be ideal.
The state she is sending is a thermal state of variance 1SNU, the rest being made up of shot noise.
This shot noise does not contain any correlation, as we have seen on Figure~\ref{cohEve}.
Yet, it brings fluctuations in the signal, so the potential for photon pairs.
This is enough to provide Alice and Bob with discord and information they can share independently of her, as we see on Figure~\ref{cohEveSN}.
This means that as soon as the eavesdropper sends a state to the legal parties, she has given them enough information that they can build a quantum secure secret key.

\begin{figure}[h]
\centering
\includegraphics[scale=0.35]{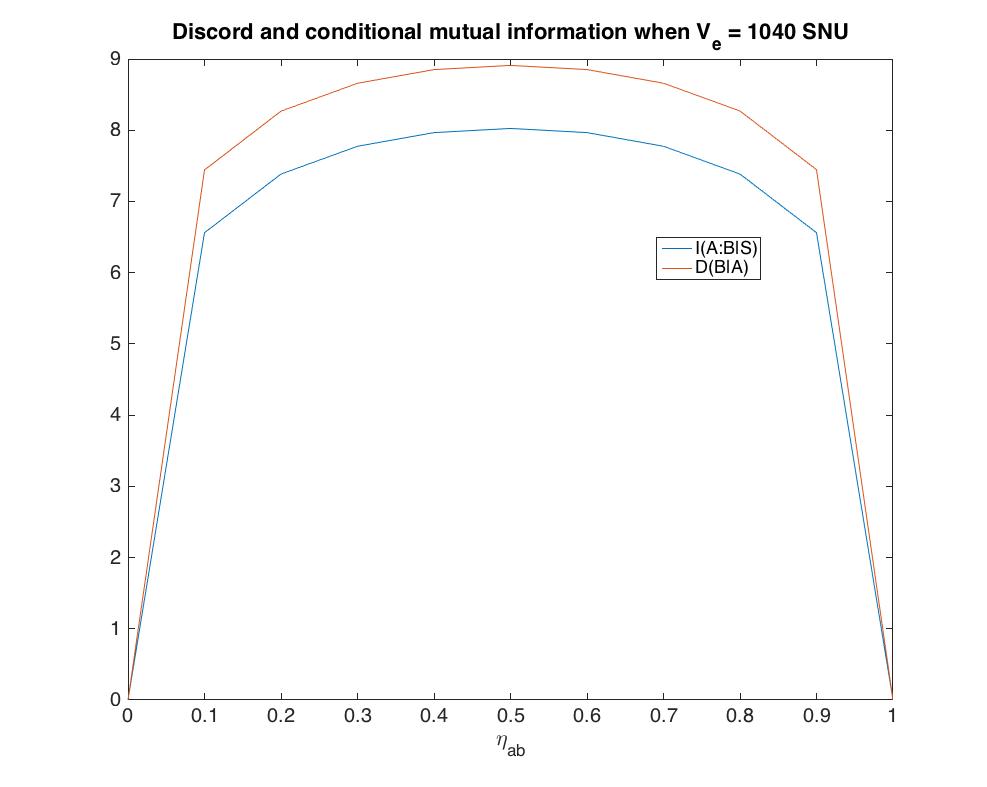}
\caption{We plot the discord (red) and conditional mutual information (blue) against $\eta_{ab}$ for $V_e = 1040SNU$. This value is consistent with an input at microwave frequencies.}
\label{cohEvethprep}
\end{figure}
Figure~\ref{cohEvethprep} illustrates how thermal preparation noise influences $I(A:B|S)$ and $D(B|A)$.
As we would naturally expect, Alice and Bob's situation is much helped by Eve's high preparation noise. 
Indeed, the higher Eve's noise at preparation, the more photon pairs arrive at $\eta_{ab}$ and become available to Alice and Bob to build a key from.

It would then be easy to conclude that a high thermal input onto $\eta_{ab}$ helps Alice and Bob, although this would perhaps be hasty.
A thermal input onto $\eta_{ab}$ can happen in essentially two ways: Eve's preparation noise is thermal (Figure~\ref{cohEvethprep}), or the channel between Eve and $\eta_{ab}$ introduces the noise. 
Commonly, the thermal channel is modelled as a beamsplitter where the ``free'' input receives a thermal state \cite{Weedbrookrmp:2012, Eisert:2005}.

\subsection{Thermal channel}
To model a thermal channel, let us add a beamsplitter of transmittance $\eta_{th}$ between the source and $\eta_{ab}$. 
At one input, we feed $V_e$ and at the other, a thermal state of variance $V_{th}$. 

The input covariance matrix to $\eta_{ab}$ is then $\Gm \bigoplus \bds{I}_2$ where  
\eqtion{\Gm = \left( \begin{array}{cccccc} \nu & 0 & \sqrt{\eta_{th}} \zt & 0   & -\mu_{th} \zt & 0 
\\ 0 & \nu  & 0  & -\sqrt{\eta_{th}} \zt & 0 &  \mu_{th} \zt 
\\ \sqrt{\eta_{th}} \zt  & 0 &  \eta_{th} \nu + \mu_{th}^2 V_{th} & 0 & \mu_{th} \sqrt{\eta_{th}} (V_{th}-\nu) & 0 
\\  0 & - \sqrt{\eta_{th}} \zt & 0 & \eta_{th} \nu + \mu_{th}^2V_{th} & 0 & \mu_{th} \sqrt{\eta_{th}} (V_{th}-\nu) 
\\ -\mu_{th} \zt & 0 & \mu_{th} \sqrt{\eta_{th}} (V_{th}-\nu) & 0 & \mu_{th}^2 \nu + \eta_{th} V_{th} & 0 
\\ 0 & \mu_{th} \zt & 0 & \mu_{th} \sqrt{\eta_{th}} (V_{th}-\nu) & 0 & \mu_{th}^2 \nu + \eta_{th} V_{th}\end{array} \right) \,.}

The output matrix becomes 
\eqtion{\Gm_{out} = \left( \begin{array}{cccc} \gm_e & \gm_{ev} & \gm_{eb} & \gm_{ea} \\ \gm_{ev} & \gm_v & \gm_{bv} & \gm_{av} \\ \gm_{eb} & \gm_{bv} & \gm_{b} & \gm_{ab} \\ \gm_{ea} & \gm_{av} & \gm_{ab} & \gm_{a} \end{array} \right)\,,}
with
 {\alld \begin{align} \gm_e =& \left( \begin{array}{cc} \nu & 0 \\ 0 & \nu \end{array} \right) \non
\\ \gm_v =& \left( \begin{array}{cc}  V_x^v & 0 \\ 0 & V_p^v \end{array} \right) \quad \text{with} \quad V_x^v = \mu_{th}^2 \nu + \eta_{th} V_{th} \quad \text{and} \quad V_p^v = \mu_{th}^2 \nu + \eta_{th} V_{th} \non
\\ \gm_a =& \left( \begin{array}{cc}  \mu_{ab}^2 V_x^{ab}  + \eta_{ab}  & 0\\ 0 & \mu_{ab}^2 V_p^{ab}  + \eta_{ab} \end{array} \right) \quad \text{with} \quad V_x^{ab} = \eta_{th}\nu + \mu_{th}^2 V_{th} \quad \text{and} \quad V_p^{ab} = \eta_{th} \nu + \mu_{th}^2 V_{th} \non
\\ \gm_b =& \left( \begin{array}{cc} \eta_{ab} V_x^{ab} + \mu_{ab}^2 & 0  \\ 0 & \eta_{ab} V_p^{ab} + \mu_{ab}^2 \end{array} \right) \non
\\ \gm_{ev} =& \left( \begin{array}{cc}  -\mu_{th} \zt & 0 \\ 0 & \mu_{th} \zt \end{array} \right) \non
\\ \gm_{eb} =& \left( \begin{array}{cc} \sqrt{\eta_{ab}}\sqrt{\eta_{th}}\zt & 0  \\ 0 & - \sqrt{\eta_{ab}}\sqrt{\eta_{th}}\zt \end{array} \right) \non
\\ \gm_{ea} =& \left( \begin{array}{cc} - \mu_{ab} \sqrt{\eta_{th}} \zt & 0 \\ 0 & \mu_{ab} \sqrt{\eta_{th}} \zt \end{array} \right) \non
\\ \gm_{bv} =& \left( \begin{array}{cc} \sqrt{\eta_{ab}}V_x^{abv} & 0 \\ 0 & \sqrt{\eta_{ab}}V_p^{abv} \end{array} \right) \quad \text{with} \quad V_x^{abv} = \mu_{th} \sqrt{\eta_{th}} (V_{th} - \nu) \quad \text{and} \quad V_p^{abv} = \mu_{th} \sqrt{\eta_{th}} (V_{th} - \nu) \non
\\ \gm_{av} =& \left( \begin{array}{cc} - \mu_{ab} V_x^{abv} & 0 \\ 0 & - \mu_{ab} V_p^{abv} \end{array} \right) \non
\\ \gm_{ab} =& \left( \begin{array}{cc} \mu_{ab} \sqrt{\eta_{ab}}(1-V_x^{ab}) & 0 \\  0 & \mu_{ab} \sqrt{\eta_{ab}}(1-V_p^{ab})  \end{array} \right) \non
\end{align}}

\begin{figure}
     \centering
     \subfloat{\includegraphics[scale=0.4]{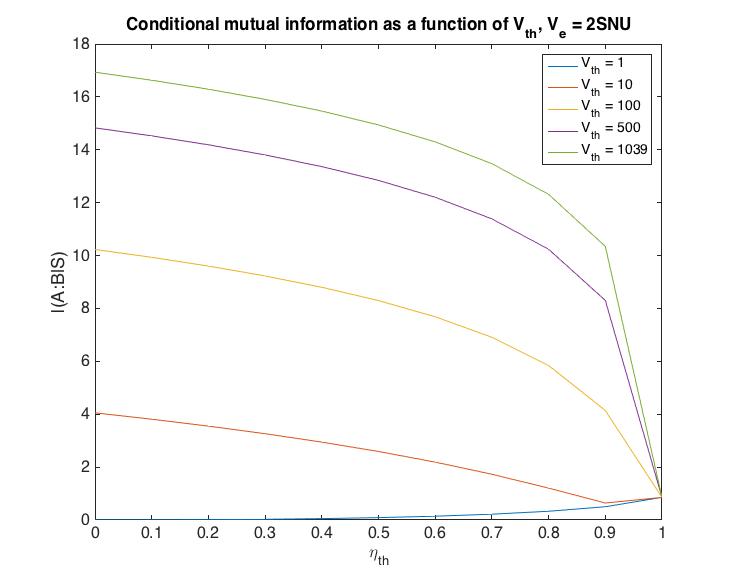} }
   \\  \subfloat{\includegraphics[scale=0.4]{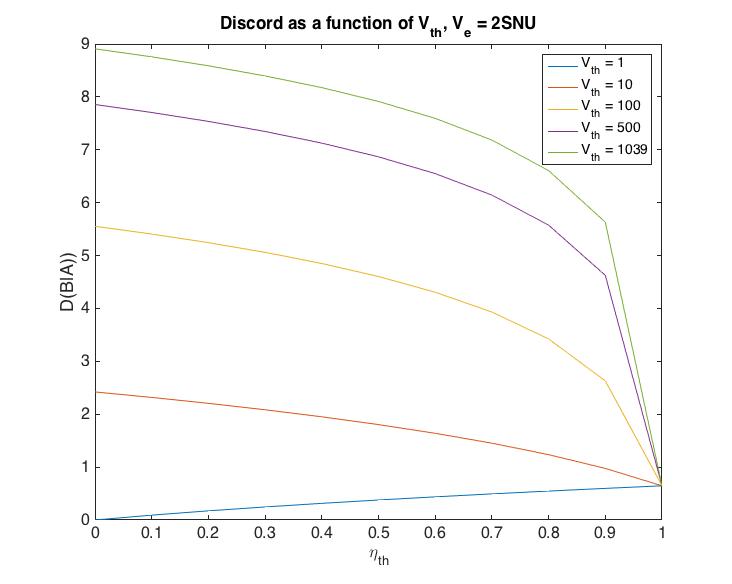} }
     \caption{Conditional mutual information and discord for several values of a thermal input at $\eta_{th}$. Eve's input is $V_e = 2 SNU$ and Alice and Bob get an equal share of the signal, $\eta_{ab} = 0.5$. }
     \label{thnoise}
\end{figure}

We can see the effect of an increasing thermal noise input on Figure~\ref{thnoise}.
When $\eta_{th} = 0$, the state which is transmitted is $V_{th}$.
As a result, when the thermal noise is very high, eg. $V_{th} = 500$SNU, $I(A:B|S)$ and $D(B|A)$ are high as $\eta_{th} \rightarrow 0$, since a large thermal state has many photon pairs to contribute to the secrecy \cite{sakuya:2018_1}.
As $\eta_{th}$ increases, Eve's contribution to the signal increases.
As $\eta_{th}$ approaches unity, $I(A:B|S)$ and $D(B|A)$ are determined by the correlations present in Eve's signal. 

The case $V_{th} = 1$SNU shows us the effect of loss in the channel.
$\eta_{th} \rightarrow 0$ models high loss and therefore, $I(A:B|S)$ and $D(B|A)$ are minimal at that point. 
This merely indicates that most of the signal is lost on the way from the satellite.
This makes the input into $\eta_{ab}$ essentially shot noise, which we have already seen, by itself has no secrecy potential.

As $V_{th}$ increase, $I(A:B|S)$ and $D(B|A)$ increase as well.
This reflects what we saw on Figure~\ref{cohEvethprep}; a higher thermal input means more photon pairs so more available information. 

Naturally, all plots converge to the same value when $\eta_{th} = 1$, since that reflects a noiseless channel. 
At that point, only the source signal provides correlations.

\section{Influence of Alice's and Bob's channel noise}

\sbp It may well seem like noise is a good thing, however, the noise considered in the previous section is noise in Eve's channel, before $\eta_{ab}$. 
Our results so far exclude channel noise in either Alice's or Bob's channels.
We now include a thermal channel between $\eta_{ab}$ and Alice, and one between $\eta_{ab}$ and Bob, in the same way as we did between the source and $\eta_{ab}$, by means of beamsplitters $\eta_{th_a}$ and $\eta_{th_b}$ with secondary thermal inputs $V_\alp$ and $V_\bt$.

The now-gigantic covariance matrix becomes 
\eqtion{\Gm = \left( \begin{array}{cccccc} \gm_e & \gm_{ev} & \gm_{eb} & \gm_{ea} & \gm_{ev_a} & \gm_{ev_b}\\ \gm_{ev} & \gm_v & \gm_{bv} & \gm_{av} & \gm_{vv_a} & \gm_{vv_b} \\ \gm_{eb} & \gm_{bv} & \gm_{b} & \gm_{ab}  & \gm_{bv_a} & \gm_{bv_b} \\ \gm_{ea} & \gm_{av} & \gm_{ab} & \gm_{a} & \gm_{av_a} & \gm_{av_b} \\ \gm_{ev_a} & \gm_{vv_a} & \gm_{bv_a} & \gm_{av_a} & \gm_{v_a} & \gm_{v_av_b} \\ \gm_{ev_b} & \gm_{vv_b} & \gm_{bv_b} & \gm_{a} & \gm_{v_av_b} & \gm_{v_b} \end{array} \right)\,,}
with relevant sub-matrices
 {\alld \begin{align} \gm_e =& \left( \begin{array}{cc} \nu & 0 \\ 0 & \nu \end{array} \right) \non
\\ \gm_a =& \left( \begin{array}{cc}  \eta_{th_a}(\mu_{ab}^2 V_x^{ab}  + \eta_{ab}) + \mu_{th_a} V_\alp ^x  & 0\\ 0 & \eta_{th_a}(\mu_{ab}^2 V_p^{ab}  + \eta_{ab}) + \mu_{th_a} V_\alp ^p \end{array} \right) \non
\\ \gm_b =& \left( \begin{array}{cc} \eta_{th_b}(\eta_{ab} V_x^{ab} + \mu_{ab}^2) + \mu_{th_b} V_\bt ^x & 0  \\ 0 & \eta_{th_b}(\eta_{ab} V_p^{ab} + \mu_{ab}^2) + \mu_{th_b} V_\bt ^p \end{array} \right) \non
\\ \gm_{eb} =& \left( \begin{array}{cc} \sqrt{\eta_{th_b}}\sqrt{\eta_{ab}}\sqrt{\eta_{th}}\zt & 0  \\ 0 & - \sqrt{\eta_{th_b}}\sqrt{\eta_{ab}}\sqrt{\eta_{th}}\zt \end{array} \right) \non
\\ \gm_{ea} =& \left( \begin{array}{cc} - \sqrt{\eta_{th_a}} \mu_{ab} \sqrt{\eta_{th}} \zt & 0 \\ 0 & \sqrt{\eta_{th_a}} \mu_{ab} \sqrt{\eta_{th}} \zt \end{array} \right) \non
\\ \gm_{ab} =& \left( \begin{array}{cc} \sqrt{\eta_{th_a}} \sqrt{\eta_{th_b}} \mu_{ab} \sqrt{\eta_{ab}}(1-V_x^{ab}) & 0 \\  0 & \sqrt{\eta_{th_a}} \sqrt{\eta_{th_b}} \mu_{ab} \sqrt{\eta_{ab}}(1-V_p^{ab})  \end{array} \right) \non
\end{align}}

\begin{figure}
\centering
\includegraphics[scale=0.45]{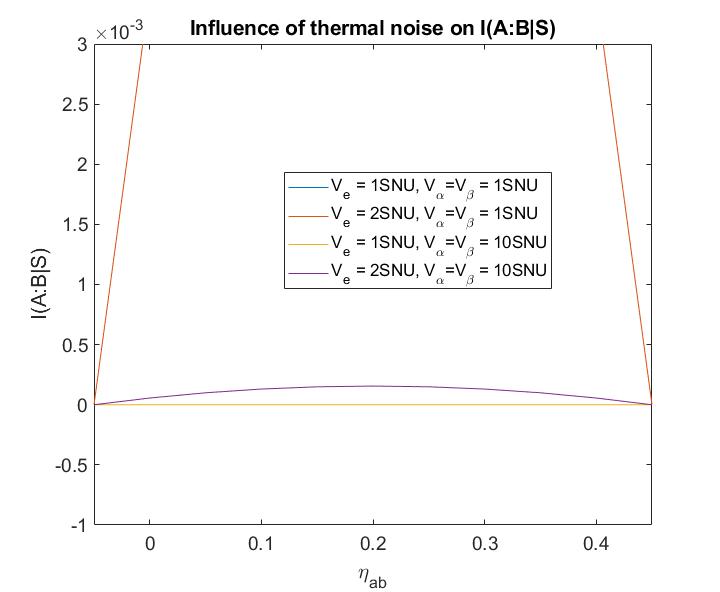}
\caption{We compare the influence of various noises on the conditional mutual information. The parameters are as follows. Blue: $V_e = 1 SNU$, $V_\alp = V_\bt = 1 SNU$. Red: $V_e = 2 SNU$, $V_\alp = V_\bt = 1 SNU$. Yellow: $V_e = 1 SNU$ $V_\alp = V_\bt = 10 SNU$. Purple: $V_e = 2SNU$, $V_\alp = V_\bt = 10 SNU$. For all plots, we consider that there is no thermal noise between the source and $\eta_{ab}$. $\eta_a = \eta_b = 0.3$.}
\label{IvsNoise}
\end{figure}

\begin{figure}
\centering
\subfloat{\includegraphics[scale=0.43]{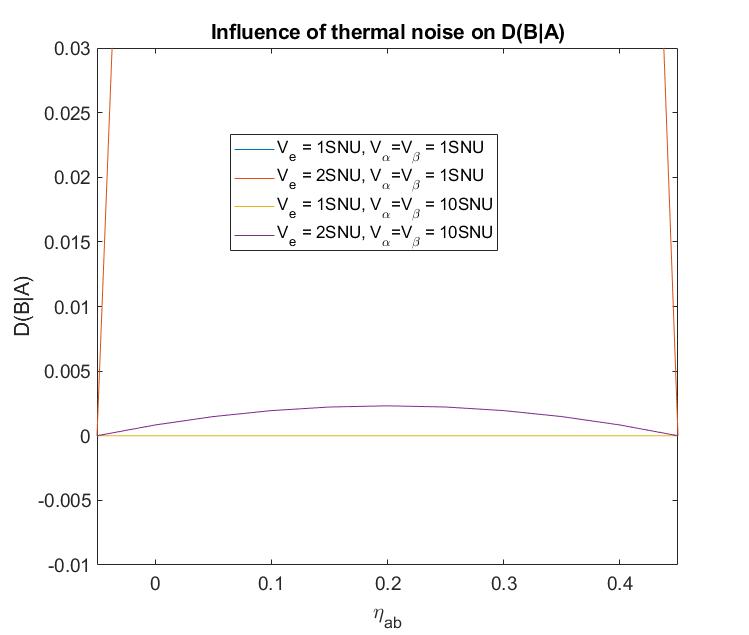} }
\caption{We compare the influence of various noises on the discord(right). The parameters are the same as for $I(A:B|S)$.}
\label{DvsNoise}
\end{figure}
Figures~\ref{IvsNoise} and \ref{DvsNoise} allow us to see the influence the thermal noise in Alice's and Bob's channels has on the conditional mutual information and on the discord.

At first glance, this seems to contradict our earlier \textit{propos}, namely, that noise is our friend. 
Indeed, we can see straight away that when Alice's and Bob's channels are noiseless and $V_e = 2SNU$ (red plot), the conditional mutual information and the discord are highest. 
This highlights that thermal noise in Alice's and Bob's channels are deleterious to secrecy. 
That is fair enough; after all, this particular noise is uncorrelated and so genuinely pollutes Alice's and Bob's respective signals, independently.

\sbp The yellow plot shows $I(A:B|S)$ and $D(B|A)$ when there is thermal noise in the legal channels (so after $\eta_{ab}$) and no source input.
In this case, there is no secrecy (and no discord) either. 
This shows the need for correlated noise, so before $\eta_{ab}$.
Indeed, the noises in Alice's and Bob's channel are not correlated. 
Their actions merely degrades the signal coming through $\eta_2$ as demonstrated by the purple curve.

\section{Closing remarks}

In this paper we have demonstrated, that any thermal source can be used for secure key exchange even if that source is held by a third party.
Alice and Bob only need verify that the statistical properties of the source are consistent with a thermal source. 
Moreover we have demonstrated throughout this paper that any noise suffered by the eavesdropper, be it preparation noise or channel noise, will benefit the legal parties. 
Furthermore, we have demonstrated this with $I(A:B|S)$, not $I(A:B)$.
This is significant. 
Not only is the mutual information between Alice and Bob positive, but the information they share independently of Eve is positive, which is a much more restrictive condition. 

Hanbury Brown and Twiss correlations can be preserved over astronomical distances and, indeed the measurement of HBT correlations forms a core component of radio astronomy.
Microwave sources are particularly convenient, because the thermal component is so high.
Indeed existing infrastructure from mobile phone masts to satellites could all be used as potential sources.

The ease of finding a source is counter balanced by the substantially reduced key rates over other methods such as CVQKD \cite{Lance:2005} or BB84 protocols \cite{Toshiba:2019}.
However, there are many low level consumer applications that do not require frequent key updates or high volumes of secret key.
It is worth noting that the aim here is not to replace or displace other methods of high speed key exchange, rather the aim is to find a protocol that would allow existing secure key exchange using existing communication mechanisms.
Indeed most modern communication systems rely on displaced thermal states (see for example \cite{Saha:1989}) and in a future paper we will aim to demonstrate key exchange using data obtained from HBT measurements in a communications system.



\sbp  The authors are grateful to network collaborators J. Rarity, S. Pirandola, C. Ottaviani, T. Spiller, N. Luktenhaus and W. Munro for very fruitful discussions, and to M. Carney and J. Wilson for help with the manuscript.
This work was supported by funding through the EPSRC Quantum Communications Hub EP/M013472/1.

\bibliographystyle{unsrt}

\bibliography{cvqkd3_bibli}



\end{document}